\documentclass[twocolumn,aps,pra,showpacs,amssymb]{revtex4}

\usepackage{graphicx}
\usepackage{amsmath}
\usepackage{amssymb}
\usepackage{bm}
\usepackage{color}

\newcommand{\bi}[1]{\mbox{\boldmath ${#1}$}}

\newcommand{\fl}{}

\newcommand{\e}{{\rm e}}
\newcommand{\rmd}{{\rm d}}
\newcommand{\rmi}{{\rm i}}

\newcommand{\up}{\uparrow}
\newcommand{\dn}{\downarrow}
\newcommand{\tr}{\hbox{tr}}

\newcommand{\half}{{\textstyle{\frac{1}{2}}}}

\newcommand{\quarter}{{\textstyle{\frac{1}{4}}}}
\newcommand{\al}{\alpha}
\newcommand{\be}{\beta}

\newcommand{\om}{\omega}
\newcommand{\Om}{\Omega}

\newcommand{\ignore}[1]{}

\def\Amp{{G_{\rm dis}}}

\def\bSigma{{\bm{\Sigma}}}

\def\half{{\textstyle \frac{1}{2} }}

\begin{document}
\title
{Suppression of non-adiabatic phases by a non-Markovian environment: \\
easier observation of Berry phases.}

\author{Robert S. Whitney}
\affiliation{
         Institut Laue-Langevin, 6 rue Jules Horowitz, B.P. 156,
         38042 Grenoble, France.}

\date{\today}
\begin{abstract}
We consider a two-level system 
coupled to a highly non-Markovian environment
when the coupling axis rotates with time.
The environment may be quantum (for example a bosonic bath or a spin bath)
or classical (such as classical noise).
We show that an Anderson 
orthogonality catastrophe suppresses transitions, so that the system's instantaneous eigenstates 
(parallel and anti-parallel to the coupling axis) can adiabatically follow
the rotation.  
These states thereby acquire Berry phases; 
geometric phases given by the area enclosed by the coupling axis.
Unlike in earlier proposals for environment-induced Berry phases,
here there is little decoherence, so one does not need a decoherence-free subspace.
Indeed we show that this Berry phase should be much easier to observe  than a conventional one, because it is not masked by either the dynamic phase or the leading non-adiabatic phase.
The effects that we discuss should be observable in any qubit device
where one can drive three parameters in the Hamiltonian
with strong man-made noise.
\end{abstract}
\maketitle


\section{Introduction}
Noise is typically a huge inconvenience
when trying to control the state of a quantum system, since
it causes dissipation and decoherence.
However one can ask if noise could be used to coherently 
control the quantum system
in a manner that 
cannot be achieve using traditional Hamiltonian manipulation.

In this context we consider
the Berry phase \cite{Berry84}; a
geometric phase associated with adiabatic evolution.  
When one rotates the parameters in a system's Hamiltonian
around a closed loop slowly enough that an eigenstate adiabatically
follows the change, then that state acquires a 
Berry phase, $\Phi_{\rm BP}$.
This phase depends on the 
geometry of the parameters' path but not on how that path is followed. 
For a spin-half in a slowly rotated magnetic field, 
$\Phi_{\rm BP} = -s_z{\cal A}$, where ${\cal A}$ is 
the solid-angle enclosed by the field and $s_z$ is the quantum
number of the spin along the field axis. 
Berry phases occur in many quantum systems 
\cite{book,BO-review,Anandan-review}, they were recently observed in
superconducting qubits \cite{Wallraff-expt}, and their noise-dependence
has been investigated using cold-neutrons \cite{Rauch-expt}.
They have potential applications in both quantum computation
\cite{Jones00Ekert00} and metrology \cite{metrology},
because it is argued that one of the most accurate ways to change the phase of
a state is to rotate the Hamiltonian's parameters round a loop which encloses 
a given solid angle.

\begin{figure}
\centerline{\includegraphics[width=7.8cm]{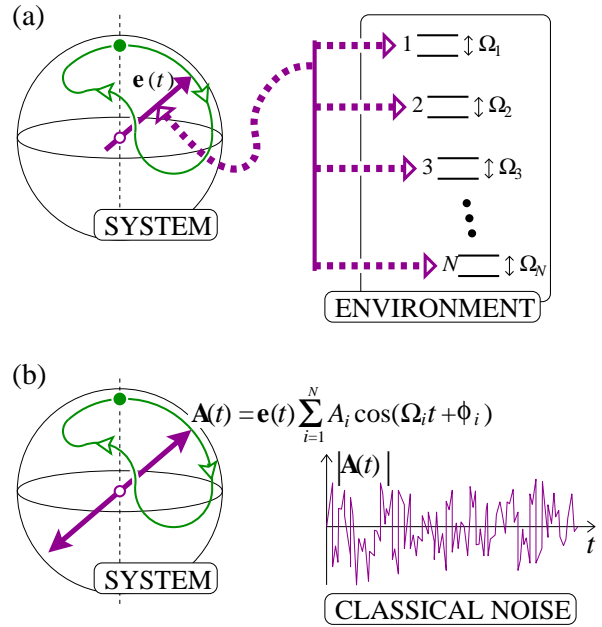}}
\caption[]{\label{Fig:qm+cl}
Sketches of the two situations we consider in this work.
(a) A Berry phase is created in the two-level system by coupling it to a quantum environment.
The environment coupling is along the axis ${\bf e}(t)$, which is 
slowly rotated around a closed loop.
(b) A Berry phase is created in the two-level system, by coupling it to
classical non-Markovian noise, ${\bf A}(t)$. 
Here the noise is along the slowly rotating ${\bf e}(t)$-axis.
In both (a) and (b), the Berry phase is given by the solid-angle enclosed by ${\bf e}(t)$.
}\end{figure}

However there is a practical problem with observing the Berry phase; 
one does not observe it ``alone''.  If the parameters of a system's 
Hamiltonian complete a closed loop
in a time $t_{\rm p}$, then the total phase acquired by an eigenstate is \cite{Berry87} 
\begin{eqnarray}
\Phi_{\rm total} 
= \Phi_{\rm dyn} + \Phi_{\rm BP} 
 +\Phi_{\rm NA}^{(1)}
+ \Phi_{\rm NA}^{(2)} + \cdots,
\label{eq:usual-total-phase}
\end{eqnarray} 
where the dynamic phase $\Phi_{\rm dyn}\propto Et_{\rm p}$,
the Berry phase $\Phi_{\rm BP}\propto (Et_{\rm p})^0$,
and the non-adiabatic (NA) correction $\Phi_{\rm NA}^{(\mu)} 
\propto (Et_{\rm p})^{-\mu}$, with $E$ being a system energy scale 
\cite{footnote:units}
(such as the gap to excitations).
As the second term in
the $t_{\rm p}^{-1}$-expansion of $\Phi_{\rm total}$,
the Berry phase is  difficult to isolate, 
and thus hard to utilize for quantum computation or metrology.
One must make $t_{\rm p}$ large to suppress the non-adiabatic terms.  
This makes $\Phi_{\rm dyn} \gg \Phi_{\rm BP}$, so 
one must subtract off $\Phi_{\rm dyn}$
(using a spin-echo trick \cite{Jones00Ekert00} or degenerate states \cite{Wilczek-Zee}) 
with extreme accuracy. 
Concretely, if one wants  $\Phi_{\rm BP}$ to an accuracy of one in $10^3$,
one requires $Et_{\rm p} \sim 10^3$. Thus $\Phi_{\rm dyn} \sim 10^3\,
\Phi_{\rm BP}$, so one must subtract off $\Phi_{\rm dyn}$
with an accuracy of one in $10^6$. Furthermore, as $t_{\rm p}$ must be long,
any device is slow to operate and leaves lots of time for
decoherence to destroy the phase.

In this work, we analyze a system coupled (via a term ${\cal H}_{\rm int}$
in the Hamiltonian) to a highly non-Markovian environment with $N$ modes 
(see Fig.~\ref{Fig:qm+cl}a).
Such environments are known to strongly renormalize the system dynamics;
inducing a type of Anderson orthogonality catastrophe \cite{Anderson},
as shown by Leggett {\it et al} using a method that they called {\it adiabatic renormalization}
\cite{Leggett,footnote:missed-by-RG}.
For large $N$, 
we show that when ${\cal H}_{\rm int}$ is changed slowly
the instantaneous eigenstates of ${\cal H}_{\rm int}$
adiabatically follow the change.
Assuming the system has no intrinsic dynamics,
the total phase acquired by an
instantaneous eigenstate
when the parameters of ${\cal H}_{\rm int}$ are 
rotated around a closed loop is
\begin{eqnarray}
\Phi_{\rm total} = \Phi_{\rm BP} + \Phi_{\rm NA}^{(2)}+ \cdots.
\label{eq:new-total-phase}
\end{eqnarray} 
Both the dynamic phase, $\Phi_{\rm dyn}$, and the first non-adiabatic phase, $\Phi_{\rm NA}^{(1)}$, are absent.
As usual the Berry phase, $\Phi_{\rm BP}$, is half the solid-angle enclosed
by the parameters of the Hamiltonian (in this case the solid-angle enclosed
by the coupling axis in ${\cal H}_{\rm int}$).

Crucially, it is much easier to accurately observe the Berry phase in a situation 
where $\Phi_{\rm total}$
is given by Eq.~(\ref{eq:new-total-phase}) in place of Eq.~(\ref{eq:usual-total-phase}).
There is no dynamic phase to subtract off, and the leading non-adiabatic phase is absent.
If we again assume we want to get the Berry phase with a accuracy of one in $10^3$,
then we require that $\Phi_{\rm NA}^{(2)} = (E_2t_{\rm p})^{-2} \sim 10^{-3}$,
which simply means that  $E_2 t_{\rm p} \sim 10^{3/2} \sim 31$
(where $E_2$ is the energy scale in $\Phi_{\rm NA}^{(2)}$).
This is a vastly less difficult to achieve than the conditions discussed above for a conventional Berry phase. 

We note that there is another geometric phase --- 
the Aharonov-Anandan phase~\cite{AA} ---
which has no non-adiabatic corrections.  Thus one may expect that one can use it
to avoid the above problems with non-adiabatic phases.  
However the Aharonov-Anandan phase has different properties from the Berry phase,
which make it more difficult to calculate in many experimental situations.  
We reserve discussion of this difficulty to Appendix~\ref{append:AA}, 
and here consider only the Berry phase.

The effects that we discuss above for a quantum environment,
are equally relevent for a system coupled to classical noise.
Classical noise is known to be equivalent to a high-temperature environment of harmonic oscillators
\cite{Caldeira-Leggett}.
We use this equivalence to show that driving a system with highly non-Markovian noise
(such as strong high-frequency noise)
whose axis is slowly rotated (see Fig.~\ref{Fig:qm+cl}b)
causes instantaneous eigenstates to
acquire a phase given by Eq.~(\ref{eq:new-total-phase}).

For either model (quantum environment or classical noise), there will
be some dephasing of the system state, however it originates from the non-adiabatic rotation of the coupling axis. 
Thus if the rotation is performed in a time $t_{\rm p}$, the dephasing goes like $\exp[-(E_1t_{\rm p})^{-1}]$ so the longer the
experiment the {\it less} the dephasing.
For all $t_{\rm p} > E_1^{-1}$, the dephasing is weak-enough to clearly measure the phase.  Thus there is no need to work in a decoherence-free subspace 
to observe
this environment-induced Berry phase (unlike that in Ref.~[\onlinecite{Carollo06}] which we discuss in Section~\ref{sect:prior-works} below).
The only difference between a low-temperature quantum 
environment and  classical noise is in this dephasing.
For low-temperature  quantum environments, 
$E_1$ grows exponentially with the strength of the environment coupling,
while for classical noise, $E_1$ goes like the squareroot of the noise power.  Thus for strong coupling to a quantum environment 
one can easily be in a situation where the $(E_1t_{\rm p})^{-1}$ is irrelevant,
then it turns out that the leading contribution to dephasing goes
like $(E_3 t_{\rm p})^{-3}$ (where we believe that $E_3$ is not exponentially suppressed), so it scales to zero much faster
with increasing $t_{\rm p}$ than for classical noise.
However we emphasize that the dephasing is already weak for classical-noise, so this fact that it is much weaker for a low-temperature quantum environment is not central message of the work we present here.
For all practical purposes, in the context of the problems we consider here, 
classical noise and quantum environments 
induce the same effects.

\subsection{Two-level system with non-Markovian quantum environment} 
\label{sect:Hs-qu}

We consider
a spin-half coupled to an environment
via a coupling axis, 
${\bf e}(t)$, 
which is rotated around a closed loop (see Fig.~\ref{Fig:qm+cl}a).
The total
Hamiltonian for the system (sys) and its environment (env) is 
\begin{subequations}\label{Eq:H}
\begin{eqnarray}
{\cal H}_{\rm sys\&env} &=& 
{\cal H}_{\rm int} (t)
+ {\cal H}_{\rm env}\big(
\{\hat a_j^\dagger,\hat a_j\}\big),
\\
{\cal H}_{\rm int}(t) &=&
-\half \hat{\bi \sigma}\cdot {\bf e}(t) \,\sum_j  K_j
(\hat a_j^\dagger +\hat a_j),
\label{Eq:Hb}
\end{eqnarray}
\end{subequations}
with 
$\hat{\bi \sigma} =
(\hat\sigma_{x},\hat\sigma_{y},\hat\sigma_{z})$
being the vector of Pauli matrices.
The unit vector ${\bf e}(t)$ is slowly rotated around a closed
loop.
The operators $\hat a_j^\dagger$ and $\hat a_j$, create and destroy
the $j$th excitation of the environment, where 
${\cal H}_{\rm env}\big(\{\hat a_j^\dagger,\hat a_j\}\big)$ is such that 
the $j$th excitation has energy $\Om_j$.
Note that this is a model of a degenerate two-level system coupled to an environment,
because we require that the two-level system's Hamiltonian is zero if one 
takes ${\cal H}_{\rm int}\to 0$ 
(we relax this requirement in Section~\ref{sect:level-splitting}).

We assume that the environment has a spectrum such that it is highly {\it non-Markovian};
by which we mean that the environment has a significant effect on 
the system's dynamics on a timescale much less than the environment's memory time, $\tau_{\rm mem}$.
This memory time is defined as the timescale on which
$\sum_j \langle \hat a_j^\dagger(\tau) \hat a_j(0)\rangle$ decays, where
$\hat a_j^\dagger(\tau), \hat a_j(\tau)$ are creation and annihilation operators in the interaction 
picture, so that 
$\hat a_j^\dagger(\tau) = \exp[{-\rmi {\cal H}_{\rm env}\tau}] \hat a_j^\dagger \exp[{\rmi {\cal H}_{\rm env}\tau}]$.

Here we say that any zero temperature environment is highly non-Markovian when the dimensionless environment-coupling parameter
\begin{eqnarray}
\Amp = \Om_{\rm m}^{-2}\sum_j K_j^2  \ \gg\ 1,
\end{eqnarray}
where $\Om_{\rm m}$ is the average frequency of the environment; defined by
$\Om_{\rm m}= \left(\sum_j K_j^2\Om_j\right)/\left(\sum_j K_j^2\right)$.
For an environment of harmonic oscillators this can be generalized to arbitrary environment temperature, $T$; such an environment is highly non-Markovian if
\begin{eqnarray}
\Amp = \Om_{\rm m}^{-2}\sum_j K_j^2\coth [\beta \Om_j/2]  \ \gg\ 1,
\end{eqnarray}
where $\beta=(k_{\rm B}T)^{-1}$.
We have in mind an environment containing so many modes with different frequencies
that the sum can be written as an integral as in Eq.~(\ref{eq:def-Amp}).

To get a feeling for such an environment, it is worth considering an environment 
at zero temperature with $N$
modes each coupled to the system with a strength $K$,
and with frequencies spread around an average of $\Om_{\rm m}$.
For simplicity we assume that the spread of frequencies is also of order $\Om_{\rm m}$,
in such a case $\Amp \simeq NK^2/\Om_{\rm m}^2$.
One can then see that  $\Amp \ll 1$ corresponds to an approximately Markovian environment
by looking at the derivation of the Bloch-Redfield master equation \cite{Bloch57,Redfield57}
which has a Markovian form 
(see for example Ref.~\onlinecite{whitney08}).  
There the memory time $\tau_{\rm mem} \sim \Om_{\rm m}^{-1}$,
while the rate at which the environment affects the system's dynamics is
$\Gamma \sim NK^2\tau_{\rm mem}$.  
The Bloch-Redfield approximation is then applicable when $\Gamma \tau_{\rm mem} \ll 1$;
then spin relaxation and dephasing times (known as $T_1$ and $T_2$ respectively) 
are of order $\Gamma^{-1}$ so $\tau_{\rm mem} \ll T_{1,2}$.
This corresponds to $\Amp \ll 1$.
In contrast the Lindblad equation master equation \cite{Lindblad,book:open-quantum-sys,whitney08}
is only applicable when $\tau_{\rm mem} \to 0$ (strictly Markovian environment)
with $NK^2$ scaled so that $\Gamma$ remains finite.  This therefore corresponds to $\Amp =0$.

In this work we consider exactly the opposite limit,   $\Amp \gg 1$.
One might guess that this corresponds to dephasing so strong that it occurs in a 
time shorter than the environment memory time. However in this regime
relaxation and dephasing rates are not given by the above $\Gamma$.
Instead there is extremely strong renormalization of the system's dynamics with 
relatively weak relaxation and dephasing.

\subsection{Environments not made of harmonic oscillators}
\label{sect:beyond-oscillators}

Most works on systems coupled to environments have considered that the environment is
made up of harmonic oscillators.  We do the same here,
treating ${\cal H}_{\rm env}= \sum_j \Om_j\hat{a}_j^\dagger \hat{a}_j$ 
(up to an irrelevant constant that we neglect).
This enables us get closed form expression for finite temperatures without too much difficulty.
However, we emphasis that the low temperature limit of our results are applicable to any large environment, for example an environment made of two-level systems.

To see why this is so, consider any large environment which is initially in its ground-state.
The analysis that follows in this article assumes that the system only weakly interacts with {\it each} degree-of-freedom in the environment, 
The effect of the environment on the system is none the less strong 
because there are so many degrees-of-freedom in the environment. 
Thus the interaction with each environment degree-of-freedom can be treated to lowest order.  This means that the interaction can only 
take environment mode $j$ from its ground state to its first excited state.
In this case the only property of environment mode $j$ that is relevant to the system's evolution 
is the gap between its ground state and its first excited state, which we define as $\Om_j$.
Thus, for temperatures such that $\e^{-\beta \Om_j}\ll 1$ for the vast majority of $j$,
the nature of environment modes is strictly irrelevant,; they could be 
harmonic oscillators (with an infinite ladder of equally spaced excited states),
spin-halves (with no states beyond the first excited state),
or any other quantum systems.

Thus one can expect that the $\beta\Om_{\rm m} \gg 1$ limit of all results in this article will apply to arbitrary environments, while the results for arbitrary $\beta$ only apply to 
an environment of harmonic oscillators.

\subsection{Two-level system with non-Markovian classical noise} 
\label{sect:Hs-cl}

One can map the dynamics of a 
system coupled to an environment of quantum oscillators
at infinite temperatures onto the ensemble-averaged dynamics of 
a system coupled to classical noise \cite{Caldeira-Leggett}.  
Thus we also analyze the problem of a spin subject to classical noise 
(with $N$ components) when the noise axis is rotated (see Fig.~\ref{Fig:qm+cl}b).
In Section~\ref{sect:classical}, we discuss what happens to a spin whose Hamiltonian is given by
\begin{eqnarray}
{\cal H}_{\rm cl} = 
-\half \hat{\bi \sigma} \cdot  {\bf e}(t)\sum_j A_j\,\cos (\Om_j t + \phi_j),
\label{Eq:Hclassicalnoise}
\end{eqnarray}
when the unit vector
${\bf e}(t)$ is slowly rotated around a closed loop,
and $A_j,\phi_j$  are random.  The amplitude $A_j$ is gaussianly distributed with 
variance $\langle A_j ^2 \rangle$, and $\phi_j$ is uniformly 
distributed over all angles from $0$ to $2\pi$.

The mapping from the quantum environment to classical noise
tells us that the average system dynamics under  ${\cal H}_{\rm cl}$
is given by taking the infinite temperature limit ($\beta=(k_{\rm B}T)^{-1} \to 0$)
of the system's dynamics under ${\cal H}_{\rm sys\& env}$
when the environment consists of harmonic oscillators and
$K_j^2$ is replaced by  $\beta \Om_j \langle A_j^2\rangle$.
Thus for such classical noise we can define the noise-coupling parameter
\begin{eqnarray}
\Amp &=& \Om_{\rm m}^{-2} \sum_n  \langle A_j^2 \rangle.
\end{eqnarray}
Here, we consider only highly non-Markovian classical noise for which $\Amp \gg 1$. 

\subsection{Prior works Berry phases with environments and 
non-adiabatic corrections.}
\label{sect:prior-works}

When one performs a Berry phase experiment on a system 
(rotating its Hamiltonian around a closed path), one can rarely ignore the
coupling to an environment (or noise).
Despite hints to the contrary  \cite{Lloyd}, 
noise-induced fluctuations of $\Phi_{\rm dyn}$ typically lead to a 
dephasing time, $T_2$,
on which all phase information (including the Berry phase) is lost 
\cite{whitney-gefen,DeChiara03,Sarandy-Lidar06}
(here we do not consider the intriguing use of repeatable noise in Ref.~[\onlinecite{Rauch-expt}]).
Thus a Berry phase can only be observed if 
one is able to adiabatically rotate the Hamiltonian 
around a closed loop in a time $t_{\rm p} \lesssim T_2$. 
Usually this requires $T_2 \gtrsim t_{\rm p} \gg E^{-1}$, 
where $E$ is the energy gap to system excitations
\cite{footnote:units}.
The environment also modifies the Berry phase \cite{whitney-gefen}; 
the modification is geometric (quadrupole-like)
and complex \cite{wmsg,wmsg2} (see also [\onlinecite{modified-BP}]),
with its imaginary part being geometric dephasing.

Ref.~[\onlinecite{Carollo06}] made the remarkable observation
that a time-dependence in the coupling to an environment 
could also {\it generate} a Berry phase 
(see also [\onlinecite{Dasgupta-Lidar07,Yuriy08}]). 
Dissipative processes (at a rate $T_2^{-1}$)
cause the system state to adiabatically follow the time-dependence
of the system-environment coupling, thereby 
acquiring a Berry phase. 
In Ref~[\onlinecite{Carollo06}], unlike in our Eq.~(\ref{eq:new-total-phase}), 
there are non-adiabatic corrections of the form 
$\Phi_{\rm NA}^{(\mu)} \sim (t_{\rm p}/T_2)^{-\mu}$ for all integer 
$\mu \geq 1$.  
To avoid the non-adiabatic effects, one requires that 
$t_{\rm p} \gg T_2$, which 
means that dephasing occurs long before the rotation is completed. 
Thus this Berry phase can
only be observed for states in a decoherence-free subspace \cite{Carollo06}.

All the works listed above considered Berry phases in systems coupled 
to environments or classical noise that were strictly or approximately Markovian.
In contrast, in this article we consider a highly non-Markovian environment, 
which strongly renormalizes the system dynamics without causing significant dissipation.
The adiabatic evolution is ensured by this renormalization 
(not dissipation), and it leads 
to Eq.~(\ref{eq:new-total-phase}). The relative lack
of dissipation means that there is no need to use a decoherence-free subspace. 

Neither the Lindblad nor Bloch-Redfield methods, 
used to study Berry phases
in Refs.~[\onlinecite{Carollo06,Dasgupta-Lidar07}] and Ref.~[\onlinecite{Yuriy08}] respectively,
can capture the strong renormalization that occurs for $\Amp \gg 1$.
The Lindblad master equation applies for $\Amp =0$, 
while the Bloch-Redfield master method applies for $\Amp \ll 1$,
which corresponds to only a very small linear renormalization effect.

We conclude by mentioning a number of works in chemical physics.
Berry phases occur in molecular dynamics because there is a separation
of timescales; the nuclei are heavy and move slowly, while the electrons
are light and move fast.  This makes it natural to perform a Born-Oppenheimer
decoupling of the fast and slow degrees-of-freedom.
It is well-known that such a de-coupling can lead to a Berry phase, 
see for example
Refs.~[\onlinecite{book,BO-review}].  There are also non-adiabatic corrections
to this Berry phase which come from violations of the Born-Oppenheimer decoupling. 
These have been well studied; for an older review see
Ref.~[\onlinecite{Yarkony96}] and references therein, 
for more recent work see Ref.~[\onlinecite{Kendrick02}].
However all the works that we are aware of, consider only a relatively small number of degrees-of-freedom (for example a tri-atomic molecule, containing three slow nuclear
degrees  of freedom and up to three fast electron degrees-of-freedom). 
In this article, we also go beyond the Born-Oppenheimer approximation
to find the non-adiabatic corrections to the Berry phase.
However we are interested in a two-level system (qubit) which is very weakly coupled to each of an enormous number of environment modes (or classical noise modes), such that the combined effect of all these modes on the system dynamics is very strong.  This is the opposite limit from that considered in the works on molecular dynamics.

\section{Berry phase due to a non-Markovian environment.}

Here we avoid using the {\it adiabatic renormalization} method of 
Leggett {\it et al} \cite{Leggett}.
Instead we perform an exact ``polaron'' transformation
on the Hamiltonian in Eq.~(\ref{Eq:Hrot}),
which fits with the spirit of  adiabatic renormalization,
while being a much more controlled approximation\cite{footnote:missed-by-RG}.
This elegant approach is standard for polarons \cite{Mahan},
and was first applied to the spin-boson model
in a number of papers \cite{Zwerger83,Silbey-Harris,Aslangul86,Dekker87,Aslangul88},
some of which are much neglected.
Refs.~[\onlinecite{Aslangul86,Dekker87,Aslangul88}] showed that the {\it non-interacting blip 
approximation} \cite{Leggett} is given by a simple weak-coupling analysis of the 
transformed Hamiltonian.
Elsewhere, we will present a detailed review of this approach 
and discuss its regime of validity (greatly over-estimated 
in Ref.~[\onlinecite{Aslangul88}]); here we simply note the remarkable
conclusion that the polaron transformation can map a 
spin coupled to a highly non-Markovian environment onto 
a spin coupled to a almost Markovian environment.
The latter can then be treated with a Bloch-Redfield master equation.

\subsection{Transforming to the rotating basis.}
\label{sect:transform-rotating}

To deal with a problem in which the axis the environment couples to is rotating with time
(as sketched in Fig.~\ref{Fig:qm+cl}), we go to a rotating basis \cite{Berry87}
whose $z$-axis remains parallel to ${\bf e}(t)$.
We transform to such a basis using
${\cal U} =
\exp(-i\half\varphi\hat\sigma_z) \exp(i\half\theta\hat\sigma_y)
\exp(i\half\varphi\hat\sigma_z)$,
in terms of polar coordinates ${\bf e}(t) = (\theta,\varphi)$.  
This choice of ${\cal U}$ gives
${\cal U} ({\bf e}(t)\cdot \hat{\bi \sigma}) {\cal U}^\dagger 
= \hat\sigma_z$ while having
no ambiguity for $\theta=0$.
In this basis, the Hamiltonian has an extra magnetic
field equal to the basis' angular velocity\cite{footnote:units}, $\bi{\om}$,
thus it is given by
\begin{eqnarray}
\label{Eq:Hrot}
{\cal H}^{\rm rot}_{\rm sys\& env} 
&=& \rmi ({\rm d}{\cal U}/{\rm d} t) {\cal U}^\dagger
+ {\cal U} {\cal H}_{\rm sys\& env} {\cal U}^{\dagger} 
\\
&=&
- {\bi{\om} \cdot \hat{\bi \sigma}\over 2} - {\hat{\sigma}_z \over 2} 
\sum_j K_j(\hat a_j^\dagger +\hat a_j) + {\cal H}_{\rm env}\big(
\{\hat a_j^\dagger,\hat a_j\}\big),
\nonumber
\end{eqnarray}
where the angular velocity in this rotating basis 
is
\begin{eqnarray}
\bi{\om}= 
\left(\begin{array}{c} \om_x \\ \om_y \\ \om_z \end{array} \right)
=
\left(
\begin{array}{c}
- \dot\varphi \sin\theta\cos\varphi - \dot\theta \sin\varphi \\
- \dot\varphi \sin\theta\sin\varphi + \dot\theta \cos\varphi \\
-\dot\varphi \,(1-\cos\theta) \end{array}
\right).
\label{eq:omega}
\end{eqnarray} 
For convenience in what follows, we define $\om_\perp$ as the magnitude of the component of $\om$ that is perpendicular to the environment-coupling axis, then 
\begin{eqnarray}
\om_\perp^2 \ =\ \om_x^2 +\om_y^2 \ =\ \dot\varphi^2 \sin^2\theta + \dot\theta^2.
\label{Eq:omega-perp}
\end{eqnarray}

Thus to summarizing the situation, we have removed the time-dependence from the coupling to the environment, by going to the rotating frame.
The Hamiltonian in Eq.~(\ref{Eq:Hrot}) is a biased spin-boson model (with the environment coupling to the spin's $z$-axis) and thus we can proceed to treat it via a polaron transformation
in a similar manner to Refs.~[\onlinecite{Zwerger83,Silbey-Harris,Aslangul86,Dekker87,Aslangul88}].

\subsection{Physics of the polaron transformation
to the basis of shifted environment modes}

\begin{figure}
\centerline{\includegraphics[width=7.5cm]{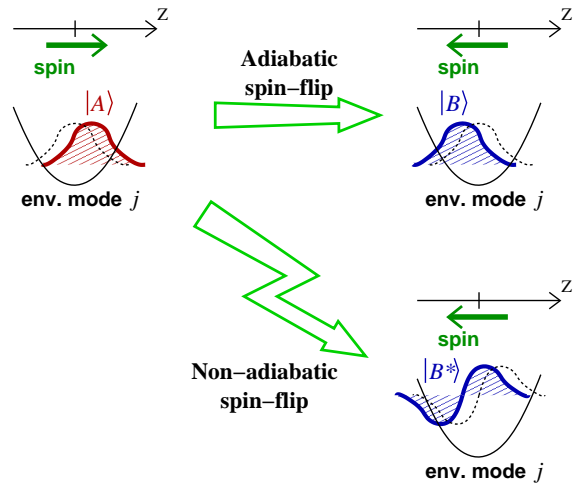}}
\caption[]{\label{Fig:orthog-cat}
A cartoon of adiabatic and non-adiabatic spin-flip transitions with
shifted environment basis-states.  Initially the spin is aligned along
the direction of the coupling axis (indicated here by $z$), the environment modes
are all shifted slightly to the right by the interaction with the spin; here we only show the $j$th mode
which is in the shifted ground-state labelled $|A\rangle$ 
(the dashed-line indicates the equivalent unshifted state).
We then show two possible spin-flip processes.
The adiabatic spin-flip, involves the $j$th mode going to state 
$|B\rangle$, which is the ground-state of the environment mode shifted slightly
to the left by the interaction with the spin.
The non-adiabatic spin-flip  involves the $j$th mode going to state 
$|B^*\rangle$, which is the excited-state of the environment mode shifted slightly
to the left by the interaction with the spin.
}\end{figure}

The physics behind the polaron transformation is the idea that {\it fast}
environment modes tend to adiabatically follow the {\it slow} system modes, so they can be thought
of as a ``cloud'' around the system, renormalizing its dynamics.
So we go to the basis of this ``cloud'', by writing 
${\cal H}_{\rm sys\& E}^{\rm rot}$
in terms of a new (orthonormal) set of basis-states 
in which environment modes
are shifted due to the force induced by the spin being $\up$ or $\dn$.
A cartoon of this (for a single environment mode) is shown in Fig.~\ref{Fig:orthog-cat}.

In this basis of shifted environment modes, any wave-function can be written as
\begin{eqnarray}
\fl |\psi \rangle &=& 
\sum_{\{m_j\}}\Big[
u_{\up, m_1,\cdots,m_N} \,
|\!\up\rangle|m_1^\up,\cdots, m_N^\up\rangle 
\nonumber \\
& & \qquad +
u_{\downarrow, m_1,\cdots,m_N} \, 
|\! \downarrow\rangle|m_1^\dn,\cdots, m_N^\dn\rangle 
\Big],
\end{eqnarray}
where  $|m_j^\up\rangle$ indicates that 
the $j$th environment mode is in the 
$m$th eigenstate of the shifted Hamiltonian
$\hat{\cal H}_j^{\rm shift;\up} = K_j(\hat{a}_j^\dagger + \hat{a}_j)/2 
+ \Omega_j\hat{a}_j^\dagger \hat{a}_j$
(we drop an irrelevant constant term).
Similarly $|m_j^\dn\rangle$ is the $m$th eigenstate of 
$\hat{\cal H}_j^{\rm shift;\dn}$, given by
$\hat{\cal H}_j^{\rm shift;\up}$ with $K_j \to -K_j$.
We assume that 
each environment mode is only weakly affected by the system, $K_j \ll \Om_j$,
then the difference between
$|m_j^\up\rangle$ and $|m_j^\dn\rangle$ is small.
Defining $\al_j \equiv (K_j/\Om_j)$, one finds,
\begin{eqnarray}
\langle m_j^\dn |m_j^\up \rangle &\simeq& 1-(m_j+1/2)\al_j^2,
\label{Eq:overlap}
\end{eqnarray}
where $\al_j=K_j/\Om_j$.
The higher the level, $m_j$, of the initial state of the $j$th 
environment mode, the smaller this overlap is. 
While this overlap is very close to one, the fact that it is slightly less than one
can have a huge effect on the system, because there are many such overlaps.
Each time the spin-flips it must carry the ``cloud'' of environment modes with it,
as a result the spin-flip rate is multiplied by the product over all $j$ of the above overlaps;
this product of $N$ overlaps --- each of which is slightly less than one ---
will decay exponentially with $N$.
Thus for large $N$, the $N$-particle environment state after the spin-flip will be almost orthogonal 
to the $N$-particle environment state before the spin-flip.
This means that the matrix element for spin-flips is suppressed (becoming exponentially small
in $N$).
Such a suppression is often called the Anderson orthogonality catastrophe \cite{Anderson},
since he pointed out that it can strongly suppress the tunnelling of an electron between two metals electrode \cite{Anderson-footnote}.

Intriguingly the higher the level, $m_j$, of the initial state of the $j$th 
environment mode, the further the overlap in Eq.~(\ref{Eq:overlap}) is from one.
This would imply that the orthogonality catastrophe is stronger at higher temperatures.
However this view is too simplistic, because it neglects possible excitations of the 
``cloud'' of environment modes that follow the spin.
The overlaps which correspond to such excitations are
\begin{eqnarray}
\fl \langle (m-1)_j^\dn |m_j^\up \rangle 
&\simeq&  m_j^{1/2}\al_j,  
\nonumber \\
\langle (m+1)_j^\dn |m_j^\up \rangle &\simeq& -(m_j+1)^{1/2}\al_j.
\label{Eq:overlap-create-annih}
\end{eqnarray}
Note that if we flip all the spins ($\up \leftrightarrow \dn$)
in  Eq.~(\ref{Eq:overlap-create-annih}),
then the right-hand-side changes sign.
We will see that these effects tend to counteract the orthogonality catatrophy which occurs in the adiabatic evolution.
Thus we must take seriously their contributions to the dynamics.

\subsection{Transforming to the basis of shifted environment-modes}

We transform Eq.~(\ref{Eq:Hrot}) to the basis of shifted environment-modes,
using a polaron transformation \cite{Mahan,Zwerger83,Silbey-Harris};
for completeness we explain the transformation in Appendix \ref{append:trans-2-shifted}.
The transformed Hamiltonian is
\begin{eqnarray}
{\cal H}^{\rm shift}_{\rm sys\& env} &=& {\cal H}'_{\rm sys} + {\cal H}'_{\rm env} 
+ {\cal V},
\nonumber \\
{\cal H}'_{\rm sys} 
&=&-\half (\omega_z \hat{\sigma}'_z + \omega_x\e^{-F} \hat{\sigma}'_x),
\nonumber \\
{\cal H}'_{\rm env} 
&=& \sum_j \Om_j \hat{a}_j^{\prime \dagger} \hat{a}'_j ,
\qquad 
\nonumber \\
{\cal V} &=& -\half \omega_x (\hat{\sigma}'_x {\cal Q}_{\rm even} 
+\hat{\sigma}'_y  {\cal Q}_{\rm odd}), 
\label{Eq:Hprime}
\end{eqnarray}
The non-interacting Hamiltonian, $({\cal H}'_{\rm sys} + {\cal H}'_{\rm env})$, involves no transitions of the environment 
modes; thus it gives adiabatic evolution. 
The interaction term, ${\cal V}$,
contains all non-adiabatic effects (i.e.  transitions between environment states).
The first term in ${\cal V}$ involves transitions
of an even (non-zero) number of environment modes,
while the second involves transitions of an odd number of environment modes.
The Hermitian operators ${\cal Q}_{\rm even,odd}$ are 
\begin{eqnarray}
{\cal Q}_{\rm even} &=& \half \e^{-F} ({\cal R}_- + {\cal R}_+),
\nonumber \\
{\cal Q}_{\rm odd} &=& {\textstyle {1 \over 2 \rmi}}\e^{-F} ({\cal R}_- - {\cal R}_+),
\label{Eq:Q}
\end{eqnarray}
where
${\cal R}_\pm = {\prod_j} 
\big[1\pm \al_j \hat{\sigma}'_z (\hat{a}_j^{\prime \dagger}-\hat{a}'_j) \big].$
The exponential suppression of perpendicular fields in ${\cal H}'_{\rm sys}$ is
given by the Franck-Condon factor,
\begin{eqnarray}
F\equiv \langle \hat{F}\rangle 
= \half \sum_j \al_j^2 \, 
\big(\, \langle\hat{a}'_j\hat{a}_j^{\prime \dagger}\rangle 
+\langle\hat{a}_j^{\prime \dagger}\hat{a}'_j\rangle\,\big).
\label{Eq:F-any-env}
\end{eqnarray} 
If the environment consists of harmonic oscillators 
at temperature $T$, then  
$F =
\half \sum_j \alpha_j^2 \,\coth \big( \beta \Om_j/2\big)$,
where $\beta=(k_{\rm B}T)^{-1}$ (see Appendix \ref{append:trans-2-shifted}).

\subsection{Berry phase from a Born-Oppenheimer approximation}
\label{sect:BO}

We can make a Born-Oppenheimer (BO) approximation
of Eq.~(\ref{Eq:Hprime}), by simply neglecting 
${\cal V}$ (since it excites environment modes).
Such a BO decoupling of
fast modes (environment) from slow modes (system) is known to create 
Berry phases \cite{BO-review}.
{\it A priori}, one might guess that this BO approximation is justified 
when the rotation-rate, $\om$, is small enough that $\om \ll \Om_{\rm m}$.
Below we find the non-adiabatic corrections and show
{\it a posteriori} that in fact the BO approximation is valid 
when $\om \ll \Om_{\rm m}\Amp^{1/2}$.
For the non-Markovian environments that we consider here ($\Amp \gg 1$),
this condition is much less stringent that the above guess.

In the BO approximation, the spin-dynamics are simply given by 
${\cal H}'_{\rm sys}$.
Assuming the Franck-Condon factor is large,
we can neglect all terms which go like $\e^{-F}$, 
then the 
Hamiltonian is simply 
$-\half \omega_z \hat\sigma'_z$.
The off-diagonal terms in this
adiabatic Hamiltonian are exponentially suppressed,
due to the Anderson orthogonality catastrophe \cite{Anderson} as outlined above.
Thus the spin adiabatically follows ${\bf e}(t)$, and the
total phase acquired by the spin is 
$ \pm \half \int_0^{t_{\rm p}} \rmd t \, \omega_z$
where $\pm$ are for $\up,\dn$.
To observe this phase, we study the precession of a superposition of  $\up$ and $\dn$,
this precession is given by the phase difference between $\up$ and $\dn$,
which is $\Phi_{\rm total} 
= \int_0^{t_{\rm p}} \rmd t \, \omega_z $
This equals the Berry phase,
\begin{eqnarray}
\Phi_{\rm BP} 
\ = \ \oint \rmd \varphi (1-\cos\theta) \ = \ {\cal A},\ 
\end{eqnarray}
where ${\cal A}$ is the solid-angle enclosed by the path of ${\bf K}(t)$.
Following Berry, we can write this as $\oint\! {\bf a}\,\rmd{\bf K}$, 
and use Stokes' theorem to give 
$\Phi_{\rm BP} = \int {\bf b}\,{\rmd{\bf S}}$ 
where ${\bf b} \equiv {\bf\nabla}_{\bf K}\times {\bf a}$
is Berry's monopole-field, $(b_K,b_\theta,b_\varphi) = 
(K^{-2},0,0)$. 

This BO analysis is sufficient to show that the spin acquires a Berry phase,
and that there is no dynamic phase, $\Phi_{\rm dyn}$.
However to see the form of the non-adiabatic phases,
we have to go beyond the BO approximation.

\section{Master equation analysis of transformed Hamiltonian}

To go beyond the BO approximation, and include the effect
of ${\cal V}$, we use a master equation approach.
The {\it exact} master equation for the evolution of the
the system's
density matrix in the shifted basis is
\begin{eqnarray}
\label{Eq:exact-master}
{\rmd \over \rmd t} \rho'_{\rm sys} &=&
 -\rmi [{\cal H}'_{\rm sys}, \rho'_{\rm sys}(t) ]_-
+\int_0^t \rmd \tau \bSigma [\tau;\rho'_{\rm sys}(t-\tau)],
\nonumber \\
\end{eqnarray}
where $\bSigma [\tau;\rho'_{\rm sys}(t-\tau)]$ 
is a super-operator 
acting on $\rho'_{\rm sys}(t-\tau)$ 
and is given by the sum of all irreducible interactions
between the system and environment, traced over all
environment modes \cite{Schoeller94,Makhlin-review03} (see also Ref.~[\onlinecite{whitney08}]).
This master equation is clearly non-Markovian, and is 
equivalent to the Nakajima-Zwanzig equation
\cite{Nakajima58-Zwanzig60}.
One gets the Bloch-Redfield equation \cite{Bloch57,Redfield57} by
treating all terms in the Eq.~(\ref{Eq:exact-master}) to second-order
in the system-environment interaction, this involves replacing  $\bSigma$
by its Born approximation $\bSigma^{\rm Born}$, while
 treating $\rho'_{\rm sys}(t-\tau)$ to zeroth order in the interaction.
Since there are effectively two 
different environments (one coupling to $\hat{\sigma}_x$ and the other to 
$\hat{\sigma}_y$) in ${\cal V}$, we identify two such second-order terms;
$\bSigma^{\rm Born}_{\rm even}$ containing a pair of ${\cal Q}_{\rm even}$-operators
and $\bSigma^{\rm Born}_{\rm odd}$ containing a pair of ${\cal Q}_{\rm odd}$-operators.
Note that cross terms (with one ${\cal Q}_{\rm even}$-  and one ${\cal Q}_{\rm odd}$-operator)
drop out when we trace over the environment because they 
contain an odd number of 
environment raising/lowering operators.
We write 
$\rho'_{\rm sys}(t-\tau) =
\e^{\rmi {\cal H}'_{\rm sys}\tau}\ \rho'_{\rm sys}(t)\ 
\e^{-\rmi {\cal H}'_{\rm sys}\tau}
= {\mathbb K}^{\rm sys}[-\tau;\rho'_{\rm sys}(t)]$,
where this defines the super-operator ${\mathbb K}^{\rm sys}$.
Then the Bloch-Redfield master equation is
\begin{eqnarray}
\fl {\rmd \over \rmd t}\rho'_{\rm sys}(t) 
&=& -\rmi [{\cal H}'_{\rm sys},\rho'_{\rm sys}(t)]_- 
+ \int_0^t \! \rmd \tau \,{\mathbb S}_{\rm even}[\tau;\rho'_{\rm sys} (t)]
\nonumber \\
& & \qquad \qquad
+ \int_0^t \! \rmd \tau \,{\mathbb S}_{\rm odd} [\tau;\rho'_{\rm sys} (t)],\quad
\label{Eq:markovian-master}
\end{eqnarray}
where
${\mathbb S}_{\rm even,odd} [\tau; X'_{\rm sys}]= 
{\bSigma}^{\rm Born}_{\rm even,odd}\big[{\mathbb K}^{\rm sys}[-\tau;X'_{\rm sys}]\big]$
with $X'_{\rm sys}$ being any system operator.
The master equation now looks Markovian (it is local in time), 
however some (weak) memory effects
are encoded~\cite{Haake-Lewenstein83,Haake-Reibold85,Suarez92} 
in ${\mathbb S}$ (see  discussion in Ref.~[\onlinecite{whitney08}]).
The super-operator ${\mathbb S}_{\rm even}$ acting on an arbitrary 
system-operator $X'_{\rm sys}$ is given by
\begin{eqnarray}
& & \hskip -8mm 
{\mathbb S}_{\rm even}[\tau;X'_{\rm sys}]
\nonumber \\
&=& 
-\quarter B_x^2 
\Big(
\big[A_{\rm even}^{\rm sym}(\tau) +\rmi A_{\rm even}^{\rm asym}(\tau) \big]
\ \hat\sigma'_x(0) \hat\sigma'_x(-\tau) X'_{\rm sys}
\nonumber \\
\fl & & \qquad 
+ \big[A_{\rm even}^{\rm sym}(\tau) -\rmi A_{\rm even}^{\rm asym}(\tau) \big]
\ X'_{\rm sys}  \hat\sigma'_x(-\tau) \hat\sigma'_x(0) 
\nonumber \\
\fl & & \qquad  - \big[A_{\rm even}^{\rm sym}(\tau) -\rmi A_{\rm even}^{\rm asym}(\tau) \big]
\ \hat\sigma'_x(0)  X'_{\rm sys}  \hat\sigma'_x(-\tau) 
\nonumber \\
\fl & & \qquad
- \big[A_{\rm even}^{\rm sym}(\tau) +\rmi A_{\rm even}^{\rm asym}(\tau) \big]
\ \hat\sigma'_x(-\tau)  X'_{\rm sys}  \hat\sigma'_x(0) \Big).
\nonumber \\
\label{Eq:boldS-even}
\end{eqnarray}
The equation for ${\mathbb S}_{\rm odd}[\tau;X'_{\rm sys}] $ is given by
Eq.~(\ref{Eq:boldS-even})
with ``even'' $\to$ ``odd'' and $\hat\sigma'_x \to \hat\sigma'_y$ throughout.
Appendix \ref{append:env-trace} gives the noise functions, 
$A_{\rm even,odd}^{\rm sym,asym}$.
We assume the environment is defined by a smooth Caldeira-Leggett $J(\Om)$-function \cite{Caldeira-Leggett}, with a typical characteristic frequency $\Om_{\rm m}$.  This function is 
defined by 
\begin{eqnarray}
\sum_j \pi K_j^2(\cdots)_{\Om_j} = \int_0^\infty \rmd \Om J(\Om)\ (\cdots)_\Om.
\label{eq:J}
\end{eqnarray}
However it is convenient to actually work with a 
dimensionless $J$-function; defined by 
$j(x) = \Om_{\rm m}^{-1} J(x\Om_{\rm m})$.
Then Appendix \ref{append:env-trace} shows that for a highly non-Markovian environment, upon neglecting terms that are of order $\exp[-2F]$ smaller than the 
leading order, one has
\begin{eqnarray}
A_{\rm even}^{\rm sym}(\tau) \! &\simeq& \!  A_{\rm odd}^{\rm sym}(\tau)  
\simeq
\half {\rm Re} \Big[\e^{- \Amp[(\Om_{\rm m}\tau)^2 +\rmi  2\chi_1 \Om_{\rm m}\tau]} \Big], \qquad
\nonumber \\
A_{\rm even}^{\rm asym}(\tau)  \! &\simeq&  \! A_{\rm odd}^{\rm asym}(\tau) 
\simeq
\half {\rm Im} \Big[\e^{- \Amp[(\Om_{\rm m}\tau)^2 +\rmi  2\chi_1 \Om_{\rm m}\tau]}\Big],
\nonumber \\
\label{Eq:all-As-strong}
\end{eqnarray}
where we define
\begin{eqnarray}
\Amp =   \int_0^\infty \rmd x\  j(x) \,  \coth[\beta\Om_{\rm m}x/2],
\label{eq:def-Amp}
\end{eqnarray}
and \begin{eqnarray}
\chi_n  \! &=&  {1 \over \Amp} \! \int_0^\infty \rmd x\  j(x) \, x^{n-2}  \left[{1 + (-1)^n\e^{-\be\Om_{\rm m}x}\over  1- \e^{-\be\Om_{\rm m}x} }\right]. \qquad
\label{Eq:chi_n}
\end{eqnarray}
The square bracket in $\chi_n$ is simply one for all odd $n$, and is $\coth[\beta\Om_{\rm m}x/2]$ 
for all even $n$. Thus by construction $\chi_2=1$. 
We note that the Franck-Condon factor, $F=\chi_0\Amp$.

\section{Non-adiabatic phases due to a quantum environment}

The phase information is contained in the off-diagonal elements of the system's density matrix.
We write this density matrix as 
$\rho'_{\rm sys} = \half (1 + s'_z \hat{\sigma}'_z) + 
s'_+ \hat{\sigma}'_+ + s'_- \hat{\sigma}'_-$
where 
$\hat{\sigma}'_\pm = (\hat{\sigma}'_x \mp \rmi \hat{\sigma}'_y)/2$ 
are the spin-raising and lowering operators.
Physically $s'_z$ is the expectation value of 
the $z'$-axis spin-polarization,
and $s'_\pm = (s'_x \pm \rmi s'_y)/2$ 
where $s'_x$ and $s'_y$ are the expectation
values of the $x'$-axis and $y'$-axis spin-polarization. 

We can rewrite Eq.~(\ref{Eq:markovian-master}) as a matrix equation 
for $s'_+$, $s'_-$ and $s'_z$, by using the fact that
$s'_\mu = \tr [\hat{\sigma}^{\prime \dagger}_\mu {\rho}'_{\rm sys}]$, for $\mu=+,-,z$.
We multiply both sides of  Eq.~(\ref{Eq:markovian-master})  by $\hat{\sigma}^{\prime \dagger}_\mu$ 
and take the trace.  To do so we evaluate the following traces
\begin{eqnarray}
\tr_{\rm sys} [
\,\hat{\sigma}^{\prime \dagger}_\mu
\,\hat{\sigma}'_x
\,\hat{\sigma}'_\nu
\,\hat{\sigma}'_x(-\tau)
\,]
\! &=&\! 
\tr_{\rm sys} [
\,\hat{\sigma}^{\prime \dagger}_\mu
\,\hat{\sigma}'_x(-\tau)
\,\hat{\sigma}'_\nu
\,\hat{\sigma}'_x
\,]
\nonumber \\
= - \tr_{\rm sys} [
\,\hat{\sigma}^{\prime \dagger}_\mu
\,\hat{\sigma}'_y 
\,\hat{\sigma}'_\nu
\,\hat{\sigma}'_y(-\tau)
\,]
\! &=&\! 
- \tr_{\rm sys} [
\,\hat{\sigma}^{\prime \dagger}_\mu
\,\hat{\sigma}'_y(-\tau) 
\,\hat{\sigma}'_\nu
\,\hat{\sigma}'_y 
\,], \qquad 
\nonumber \\
\label{Eq:traces-that-cancel}
\end{eqnarray}
and 
\begin{eqnarray}
\tr_{\rm sys} [
\, \hat{\sigma}^{\prime \dagger}_\mu
\,\hat{\sigma}'_x 
\,\hat{\sigma}'_x(-\tau)
\,\hat{\sigma}'_\nu
\,]
=
\tr_{\rm sys} [
\,\hat{\sigma}^{\prime \dagger}_\mu
\,\hat{\sigma}'_\nu
\,\hat{\sigma}'_x(-\tau)
\,\hat{\sigma}'_x
\,]
\nonumber \\
\quad = \tr_{\rm sys} [
\,\hat{\sigma}^{\prime \dagger}_\mu
\,\hat{\sigma}'_y 
\,\hat{\sigma}'_y(-\tau)
\,\hat{\sigma}'_\nu
\,]
=
\tr_{\rm sys} [
\,\hat{\sigma}^{\prime \dagger}_\mu
\,\hat{\sigma}'_\nu 
\,\hat{\sigma}'_y(-\tau) 
\,\hat{\sigma}'_y
\,]
\nonumber \\
=
\left(
\begin{array}{ccc}
\Xi & 0 & 0 \\
0 & \Xi^* & 0 \\
0 & 0 & 2{\rm Re}[\Xi]
\end{array} \right)_{\mu\nu},
\end{eqnarray}
where for compactness we define $\Xi \equiv \e^{-\rmi \om_z \tau}$.
We do not need the matrix form of the traces in Eq.~(\ref{Eq:traces-that-cancel}),
 because the presence of the minus signs causes the terms to {\it cancel} amongst themselves,
so they play no further role in our analysis.

Now we note that the matrix equation for $s'_+$, $s'_-$ and $s'_z$, is actually
three uncoupled equations.  Thus to see the phase information we need only analysis
the equation for $s'_+$, which reads
\begin{eqnarray}
{\rmd \over \rmd t} s'_+ = \left[-\rmi \om_z  - \half \om_\perp^2
\int_0^\infty  \rmd \tau A_{\rm even}^{\rm sym}(\tau) \e^{-\rmi \om_z \tau} \right] s'_+\, ,
\label{Eq:diff-eqn-s_+}
\end{eqnarray}
where $\om_\perp$ is given in Eq.~(\ref{Eq:omega-perp}),
and $A_{\rm even}^{\rm sym}(\tau)$ is given by Eq.~(\ref{Eq:all-As-strong}).
The integral over $\tau$  can be written as
\begin{eqnarray}
& & \hskip -10mm \int_0^\infty  \rmd \tau A_{\rm even}^{\rm sym}(\tau) \e^{-\rmi \om_z \tau} 
\nonumber \\
&=&
{1 \over 4}
\left [ I \left({\om_z\over 2\Amp\Om_{\rm m}}\right) 
 + I^*\left({-\om_z \over 2\Amp\Om_{\rm m}} \right) \right],
\label{Eq:A-integral}
\end{eqnarray} 
where we define
\begin{eqnarray}
I(b) \! &\equiv&\! \int_0^\infty \! \! \rmd \tau \,
\exp \Big[-\half \Amp 
\Big((\Om_{\rm m}\tau)^2+\rmi 2(\chi_1+ b)\Om_{\rm m}\tau\Big)\Big].
\nonumber \\
\end{eqnarray}
The real part of $I(b)$ can found exactly by writing the integral as one from $-\infty$ to $\infty$ and completing the square, the result is exponentially suppressed for large $\Amp\chi_1^2$,
\begin{eqnarray}
{\rm Re}[I(b)] &=& {\pi^{1/2} \e^{- \Amp (\chi_1+ b)^2/2 } \over \Om_{\rm m} (\Amp/2)^{1/2}}.
\label{Eq:Iresult-real}
\end{eqnarray}
In contrast the imaginary part has no exponential suppression with $\Amp\chi_1^2$. For large $\Amp\chi_1$ we can drop the quadratic term in the exponent, and find
\begin{eqnarray}
{\rm Im}[I(b)] &=& 
- { 1 \over \Om_{\rm m} \Amp (\chi_1+ b)} \big(1 +{\cal O}[(\Amp\chi_1)^{-1}] \big). \qquad
\label{Eq:Iresult-imag}
\end{eqnarray}
Now we use 
Eqs.~(\ref{Eq:Iresult-real},\ref{Eq:Iresult-imag}) to expand Eq.~(\ref{Eq:A-integral})  in powers of $\om_z$.
Then we cast the right hand side of Eq.~(\ref{Eq:diff-eqn-s_+}) in terms of powers of $\om_{\perp,z}$, so it reads
\begin{eqnarray}
{\rmd \over \rmd t} s'_+ &=& \Big[
-\rmi \om_z  - \gamma_1\e^{-\Amp\chi_1^2/2}\, \om_\perp^2  +\rmi \gamma_2 \om_\perp^2 \om_z 
\nonumber \\
& & \qquad \qquad \qquad \qquad
+ {\cal O}[\gamma_1^2]\, \om_\perp^3
 \Big] s'_+ \, ,
\end{eqnarray}
where 
\begin{eqnarray}
\gamma_1= \Om_{\rm m}^{-1}[2\pi/\Amp]^{1/2}, \qquad 
\gamma_2 = (2\Om_{\rm m} \Amp\chi_1)^{-2}.
\label{Eq:gammas}
\end{eqnarray}
The $\om_\perp^3$-term is beyond Bloch-Redfield method applied in this work.  We will discuss it in detail elsewhere, here we simply note that dimensional analysis
shows that it goes like $\gamma_1^2$ (it is a third order contribution to ${\mathbb S}_{\rm even,odd}$
and thus contains a double time-integral whose integrand is dominated by times of order $\gamma_1$).
It is trivial to solve the above equation for $s'_+$, 
the solution is
\begin{eqnarray}
s'_+ (t_{\rm p}) &=& \exp[-\rmi \Phi_{\rm total}-D_{\rm total}\, ] \,s'_+ (0) ,
\end{eqnarray}
where the phase $\Phi_{\rm total}$ and the dephasing factor $D_{\rm total}$
are given by
\begin{eqnarray}
\Phi_{\rm total} \!\!&=&\! \! \int_0^{t_{\rm p}}\!\! \rmd t \Big[\om_z
- \gamma_2 \,\om_\perp^2 \om_z + {\cal O}[\gamma_1^2] \,\om_\perp^3 \,\Big],
\label{Eq:Phi-total-finalresult}
\\
D_{\rm total} \!\!&=&\! \! \int_0^{t_{\rm p}}\!\! \rmd t \Big[ \gamma_1\e^{-\Amp\chi_1^2/2}\om_\perp^2 + {\cal O}[\om_\perp^4] \,\Big].
\label{Eq:D-total}
\end{eqnarray}
Noting that $\om_{x,z}$ go like $t_{\rm p}^{-1}$,
we immediately get
\begin{subequations}\label{Eq:all-phases}
\begin{eqnarray}
\Phi_{\rm dyn} &=&0,
\\
\Phi_{\rm BP}&=&\int_0^{t_{\rm p}} \rmd t \, \om_z \, =\, {\cal A},
\\
\Phi_{\rm NA}^{(1)}  &=&  0,
\label{eq:Phi_NA1}
\\
\Phi_{\rm NA}^{(2)} &=& - \int_0^{t_{\rm p}} \rmd t \,
\left( \gamma_2 \,\om_\perp^2 \om_z + {\cal O}[\gamma_1^2]\, \om_\perp^3 \right),
\end{eqnarray}
\end{subequations}
where ${\cal A}$ is the solid-angle enclosed by the coupling-axis (see Section~\ref{sect:BO}).
Higher-order non-adiabatic phases will appear, however just like the ${\cal O}[\om_\perp^3]$ term in the integrand of  $\Phi_{\rm NA}^{(2)}$, most of them are beyond the Born approximation used to get the Bloch-Redfield equation, Eq.~(\ref{Eq:markovian-master}).

We draw attention to the fact that $\Phi_{\rm NA}^{(1)}$ is zero.
The term in $(-\rmi \Phi_{\rm total}-D_{\rm total})$ which goes like $t_{\rm p}$ is purely real; thus it contributes to the total dephasing factor, 
$D_{\rm total}$, but not the total phase, $\Phi_{\rm total}$.

Now we turn to the total dephasing factor $D_{\rm total}$,
 we split it in to dynamic, geometric,
and non-adiabatic contributions (just as we did for $\Phi_{\rm total}$). 
We define  {\it dynamic dephasing} as terms in  $D_{\rm total}$
that go linearly with time (thus the usual $\exp[-t/T_2]$ dephasing 
of two-level systems by weak-Markovian noise is dynamic dephasing).
We define {\it geometric dephasing} (a term coined in Ref.~[\onlinecite{wmsg}]) as terms in  $D_{\rm total}$
that are $t_{\rm p}$-independent, while 
we define {\it non-adiabatic corrections to dephasing}
as terms that go like $t_{\rm p}$ to some negative power.
Then we immediately see that
$D_{\rm dyn} = D_{\rm BP} = D_{\rm NA}^{(2)}=0$, 
while
\begin{eqnarray}
D_{\rm NA}^{(1)} &=& \int_0^{t_{\rm p}}\!\! \rmd t \  \gamma_1\e^{-\Amp\chi_1^2/2}\om_\perp^2. 
\label{Eq:D_NA1}
\end{eqnarray}
There will also be a finite $D_{\rm NA}^{(3)}$ term beyond the Bloch-Redfield analysis.

We assumed from the beginning that the Franck-Condon factor, $F=\Amp\chi_0$ is sufficiently large that we can neglect those exponentially small effects that go like $\e^{-F}$.  Thus for a finite temperature
environment it is  natural to say 
that $\Amp \chi_1^2$ is also sufficiently large that
$D^{(1)}_{\rm NA}\propto \e^{-\Amp \chi_1^2/2}$ is very small.
Although this is not the case at very high temperature, 
since then $\chi_1 \to 0$. 
However for any finite temperature,  
as one increases the coupling, $\Amp$,
one will rapidly reach the situation where $D^{(1)}_{\rm NA}$ is so small that dephasing is given by $D_{\rm NA}^{(3)} \sim t_{\rm p}^{-3}$
(which we believe does not have a similarexponential suppression).
Then we can effectively neglect all dephasing in the situation
where $t_{\rm p}$ is long enough for accurate observation of the
Berry phase. By this we mean that when $t_{\rm p}$ is long enough to 
ensure that $\Phi_{\rm NA}^{(2)} \ll \Phi_{\rm BP}\sim 1$, one will also have 
$D_{\rm total} \sim D_{\rm NA}^{(3)} \ll \Phi_{\rm NA}^{(2)} \ll 1$,
so $D_{\rm total}$ will be tiny.

With the aid of the definition of $\gamma_1$ and $\Amp$, we find
\begin{eqnarray}
\Phi_{\rm NA}^{(2)} \sim \left( {1 \over NK^2} {\Om_{\rm m}^2 \over NK^2} + {\cal O}\left[(NK^2)^{-1}\right]\right) {1 \over t_{\rm p}^2},
\end{eqnarray}
where $N$ is the number of environment modes and $K$ is the coupling to each one
(so $NK^2$ gives the strength of the coupling to the environment).
In the derivation we assumed strong dissipation, $\Amp \sim NK^2/\Om_{\rm m}^2 \gg 1$,
thus the second term (which comes from the $\omega_x^3$-term) dominates.
Unfortunately this is the term whose exact form we do not have.  None the less we see
it is sufficient to make the $NK^2$ as large as possible to minimize 
$\Phi_{\rm NA}^{(2)}$.

\section{Non-adiabatic phases due to a classical noise-field.}
\label{sect:classical}

Here we consider the situation sketched in Fig.~\ref{Fig:qm+cl}b,
and discussed in Section \ref{sect:Hs-cl}.
As shown in Ref.~[\onlinecite{Caldeira-Leggett}], this problem is equivalent to the above quantum problem for an environment of harmonic oscillators,
with $K_j^2$ replaced by $\beta\Om_j \langle A_j^2\rangle$, in the infinite
temperature limit $\beta = (k_{\rm B}T)^{-1} = 0$.
If we assume there is a continuum of frequencies in the classical noise,
it is natural to define
\begin{eqnarray}
\sum_j \pi \langle A_j^2\rangle \, (\cdots)_j 
= \int_0^\infty \rmd \Om J_{\rm cl}(\Om) \, (\cdots)_\Om\, ,
\end{eqnarray}
the factor of $\pi$ is to make the analogy with Eq.~(\ref{eq:J}).
Then $J_{\rm cl}(\Om)$ can be thought of as a measure of the noise power at 
frequency $\Om$.
Once again we define a dimensionless J-function;
$j_{\rm cl}(x) = \Om^{-1}_{\rm m} J_{\rm cl}(x\Om_{m})$.
Then we can get results for classical noise from the results we have for a quantum environment;
we simply need to replace
$j(x)$ in all the formulae 
in the preceding section of this work with $\beta\Om_{\rm m}\, x\, j_{\rm cl}(x)$ 
and then take the limit $\beta \to 0$.
Then we find that
\begin{eqnarray}
\Amp &=& \int_0^\infty  \rmd x \ j_{\rm cl}(x),
\end{eqnarray}
and 
\begin{eqnarray}
\hbox{Even  }n:  \  \chi_n \! &=&\!  {1\over \Amp}
\int_0^\infty{ \rmd x \ j_{\rm cl}(x)\over x^{2-n}},
\\
\hbox{Odd  }n: \  \chi_n \! &=& \lim_{\be =0} \left[{\beta\Om_{\rm m} \over \Amp}\! 
\int_0^\infty {\rmd x \ j_{\rm cl}(x)\over x^{1-n}}\right] \longrightarrow 0.
\qquad 
\end{eqnarray}
In this situation, we see that Franck-Condon factor, $F=\Amp\chi_0$, remains finite
as $T \to \infty$.  Thus if we have strong non-Markovian noise, i.e.~large $\Amp$, we can have large $F$ for which the BO analysis in section~\ref{sect:BO} 
shows us that rotating classical noise will induce a Berry phase in the spin.  
However for such classical noise, $\Amp \chi_1\to 0$,
so there will be no exponential suppression of ${\rm Re}[I(b)]$ given in Eq.~(\ref{Eq:Iresult-real}).
In addition we cannot use the result for ${\rm Im}[I(b)]$ given in Eq.~(\ref{Eq:Iresult-imag}),
because it is only for large $\Amp \chi_1$.
Instead we note that for $\Amp \chi_1\to 0$, one has
\begin{eqnarray}
 \lim_{\Amp \chi_1= 0}{\rm Im}[I(b)]
&=&  {2b \over \Om_{\rm m}\Amp}  +{\cal O}[b^3].
\end{eqnarray}
Hence $\gamma_1$  remains the same as in Eq.~(\ref{Eq:gammas}), but now 
\begin{eqnarray}
\gamma_2 = [2\Om_{\rm m}^2 \Amp^2]^{-1}.
\label{Eq:gamma_2class}
\end{eqnarray}
From this we find that the total phase, $\Phi_{\rm total}$, is that given by
Eq.~(\ref{Eq:Phi-total-finalresult}). Thus all phases
(the dynamic phase, Berry phase,
and non-adiabatic phases) are basically 
the same for the classical 
noise as they
were for the quantum environment. They are given by Eq.~(\ref{Eq:all-phases}) with the only difference being that now 
$\gamma_2$ is given by Eq.~(\ref{Eq:gamma_2class}).

 Turning to the dephasing factor, $D_{\rm total}$, 
the fact that ${\rm Re}[I(b)]$ is given by Eq.~(\ref{Eq:Iresult-real}) 
with $\Amp \chi_1=0$
means that in place of Eq.~(\ref{Eq:D_NA1}) we have 
\begin{eqnarray}
D^{(1)}_{\rm NA} &=& \int_0^{t_{\rm p}} \rmd t \ \gamma_1 \, \omega_x^2.
\end{eqnarray}
We still have $D_{\rm dyn}=D_{\rm BP}=D_{\rm NA}^{(2)}=0$.
Since now $D^{(1)}_{\rm NA}$ has no exponential suppression, it will dominate the dephasing, which will be non-negligible.
To observe the Berry phase in the presence of this dephasing, 
one requires that 
$D^{(1)}_{\rm NA} \lesssim 1$.
Taking $\om_\perp \sim t_{\rm p}^{-1}$  we have 
\begin{eqnarray}
D^{(1)}_{\rm NA}\  \sim \ \left[ {\textstyle \int_0^\infty} \rmd \Om \, J_{\rm cl} (\Om) \right]^{-1/2} t_{\rm p}^{-1} 
\ \sim \ \big[\sqrt{NA^2} \ t_{\rm p} \big]^{-1}. \quad
\end{eqnarray}
The right-hand side contains the integrated noise power,
which is of order $NA^2$ for noise with  $N$ modes each with an amplitude $A$.
This means that the longer the evolution takes, 
the less dephasing there will be.
Once $t_{\rm p} \gtrsim (NA^2)^{-1/2}$, we will be able to measure the Berry phase without worrying about dephasing.

The leading non-adiabatic correction to the phase is $\Phi_{\rm NA}^{(2)}$,
it goes like 
\begin{eqnarray}
\Phi_{\rm NA}^{(2)} \sim \left( {1 \over NA^2} {\Om_{\rm m}^2 \over NA^2} + {\cal O}\left[(NA^2)^{-1}\right]\right) {1 \over t_{\rm p}^2}.
\end{eqnarray}
This is much the same as for the quantum environment (despite the different form for $\gamma_2$),
once again it is dominated by the second term which comes from the $\omega_x^3$-term.
Thus to have small  $\Phi_{\rm NA}^{(2)}$ one needs large $NA^2$.

These conditions are less strict than the equivalent 
one for a conventional Berry phase observation. 
Since the dephasing is weak,
the accuracy with which one can observe the Berry phase,
is determined by how small $\Phi_{\rm NA}^{(2)}$ is.
The total phase will equal the Berry phase to within 0.1\% ,
whenever $\Phi_{\rm NA}^{(2)} \lesssim 10^{-3}\Phi_{\rm BP}$,
this requires that the noise is strong enough that 
$N^{1/2}A \,t_{\rm p} \gtrsim 10^{3/2} \sim 31$.
In this situation $D_{\rm total} \sim D_{\rm NA}^{(1)} \sim 10^{-3/2}$, so 
$\exp[-D_{\rm total}] \sim 0.97$, which means that dephasing is indeed
extremely weak; it only suppresses the signal is only about 3\%.

\section{Finite level-splitting in the qubit}
\label{sect:level-splitting}

Thus far, we have assumed that ${\cal H}_{\rm sys \& env}$
does not contains a ${\cal H}_{\rm sys}$-term,
where a ${\cal H}_{\rm sys}$-term would be one that contains system-operators but not environment operators.
Physically this means that the qubit levels would be degenerate if we 
turned-off the environment coupling
in Eq.~(\ref{Eq:H}) or turned-off the noise in Eq.~(\ref{Eq:Hclassicalnoise}).
However it is natural to ask what happens to the effects that we discuss
in a system where this degeneracy is not perfect.
Indeed, this may be important for observation in qubits, where one must tune a succession of external gate voltages and fields until the qubit is
as close to  degeneracy as possible.  This means that we
can assume  that ${\cal H}_{\rm sys}$ is small, but not exactly zero
(one only has an exact degeneracy if it is due to some symmetry of the system).

Thus we now briefly revisit the calculations that we performed throughout this work, adding a small $H_{\rm sys}$-term.
This means the Hamiltonian is
\begin{eqnarray}
{\cal H}_{\rm sys\&env} &=& {\cal H}_{\rm sys}(t) +
{\cal H}_{\rm int} (t)
+ {\cal H}_{\rm env}\big(
\{\hat a_j^\dagger,\hat a_j\}\big),
\nonumber
\\
{\cal H}_{\rm sys}(t) &=&
-\half \hat{\bi \sigma}\cdot {\bf B}(t),
\end{eqnarray}
with ${\cal H}_{\rm int}$ given by Eq.~(\ref{Eq:Hb}).
If we now do exactly the same transformation to go the rotating frame
as in Section~\ref{sect:transform-rotating}.
We see that the resulting Hamiltonian is the same as
that given in Eq.~(\ref{Eq:Hrot}) with the following substitutions
\begin{eqnarray}
\omega_x \to \omega_x + B_x\, ,
\nonumber \\ 
\omega_y \to \omega_y + B_y\, ,
\\
\omega_z \to \omega_z + B_z\, ,
\nonumber 
\end{eqnarray}
where $B_x,B_y,B_z$ are the components of ${\bf B}(t)$ written in the
rotating frame, so $B_z$ is the component of ${\bf B}(t)$ parallel
to the environmental coupling axis (defined by ${\bf e}(t)$).
Thus following through the calculation, we find that
Eqs.~(\ref{Eq:Phi-total-finalresult},\ref{Eq:D-total}) are replaced by
\begin{eqnarray}
\Phi_{\rm total} &=& \int_0^{t_{\rm p}}\!\! \rmd t \Big[\,\om_z+B_z 
\nonumber \\
& & \quad - \gamma_2 \,
\big[ (\om_x+B_x)^2+(\om_y+B_y)^2\big][\om_z +B_z]
\nonumber \\
& & \quad + {\cal O}[\gamma_1^2] \,
\big[ (\om_x+B_x)^2+(\om_y+B_y)^2\big]^{3/2} \Big],
\\
D_{\rm total} &=& \int_0^{t_{\rm p}}\!\! \rmd t 
\Big[\,  \gamma_1 \e^{-\Amp\chi_1^2/2}
\big[ (\om_x+B_x)^2+(\om_y+B_y)^2\big]
\nonumber \\
& & \qquad \qquad \qquad
+ {\cal O}[(\om+B)^4] \Big].
\end{eqnarray}
Now noting that
$\om_{x,y,z}$ goes like $t_{\rm p}^{-1}$, while $B_{x,y,z}$ is independent of $t_{\rm p}$, we can
expand in powers of  $t_{\rm p}^{-1}$.
Then we find the phases
\begin{subequations}\label{Eq:phases-non-degen}
\begin{eqnarray}
\Phi_{\rm dyn} &=& \int_0^{t_{\rm p}}\!\! \rmd t   \, B_z\, ,
\\
\Phi_{\rm BP} &=&  \int_0^{t_{\rm p}}\!\! \rmd t \Big[\,\om_z 
+ {\cal O}[\gamma_2 B^2 \om,\gamma_1^2 B^2\om] \,\Big],
\label{Eq:BP-non-degen}
\\
\Phi_{\rm NA}^{(1)} 
&=&   {\cal O}[\gamma_2 B ,\gamma_1^2 B] \,  t_{\rm p}^{-1},
\label{Eq:NA1-non-degen}
\\
\Phi_{\rm NA}^{(2)} 
&=&   {\cal O}[\gamma_2 ,\gamma_1^2] \,  t_{\rm p}^{-2},
\end{eqnarray}
\end{subequations}
and dephasing factors
\begin{subequations}\label{Eq:Ds-non-degen}
\begin{eqnarray}
D_{\rm dyn} &=&  \int_0^{t_{\rm p}}\!\! \rmd t 
\  \gamma_1 \e^{-\Amp\chi_1^2/2} \big[ B_x^2+B_y^2\big],
\\
D_{\rm BP} &=& \int_0^{t_{\rm p}}\!\! \rmd t \ 
2\gamma_1 \e^{-\Amp\chi_1^2/2} (B_x\om_x + B_y\om_y),  \qquad
\label{Eq:D_BP-non-degen}
\\
D_{\rm NA}^{(1)} &=& \int_0^{t_{\rm p}}\!\! \rmd t \ 
 \gamma_1\e^{-\Amp\chi_1^2/2} (\om_x^2+\om_y^2).
\end{eqnarray}
\end{subequations}

Our first observation is that the system develops a dynamic phase,
$\Phi_{\rm dyn}$ which is proportional to $Bt_{\rm p}$.
Thus if one wishes to observe the Berry phase alone to high accuracy, one should ensure that $\Phi_{\rm dyn}$ is sufficiently small; for example,
for an accuracy of 1 in $10^3$ we need $\Phi_{\rm dyn} < 10^{-3}$.
This provides a simple criteria for  $B t_{\rm p}$,
which correspondingly say how close to degenerate the states should be
in the absence of the environmental coupling or classical noise 
The requirement that $Bt_{\rm p}\ll 1$ corresponds to $B \ll \om$,
thus the higher the power of $B$ in Eqs.~(\ref{Eq:phases-non-degen},
\ref{Eq:Ds-non-degen}), 
the smaller the contribution.

Turning to dephasing, we see that there are now small
dynamic dephasing and geometric dephasing terms.
For finite-temperature quantum environments, 
these are all exponentially suppressed for strong environment coupling,
but for classical noise ($\chi_1=0$) they are only powerlaw suppressed.
However one should not forget that we are interested in the 
regime where $B \ll \om$ and hence 
$D_{\rm dyn} \ll D_{\rm BP} \ll D_{\rm NA}^{(1)}$.
Thus the dephasing will be dominated by $D_{\rm NA}^{(1)}$
in situations where $D_{\rm NA}^{(1)}$ is not exponentially small, and 
be dominated by $D_{\rm NA}^{(3)}$
in situations where  $D_{\rm NA}^{(1)}$ is exponentially small.
We note that there is finite geometric dephasing here,
therefore we should add this system (one coupled to strong non-Markovian noise) to the list of situations in which geometric dephasing occurs
(non-Hermitian Hamiltonians~\cite{Garrison88}, weak Markovian noise \cite{wmsg},
or ultra-slow noise \cite{other-geometric-dephasing}).
However in each of these situations, the form of the geometric dephasing
is rather different.

The presence of finite $B$ gives a finite $\Phi_{\rm dyn}$, thus one might
ask if it is still easier to observe this Berry phase than a 
conventional Berry phase.
The answer is yes.  In the conventional Berry phase the same energy
parameter, $E$, that controls the smallness of non-adiabatic phase $\propto (Et_{\rm p})^{-1}$, enters the dynamic phase $\propto Et_{\rm p}$.
Thus choosing $E$ to minimize the former, maximizes the latter.
Here non-adiabatic corrections are controlled by one parameter (the
strength of coupling to the environment, $\Amp$), 
while the dynamic phase is controlled
by another (the magnitude of the field, $B$).  
Thus both these unwanted phases can be independently minimized 
(by taking large $\Amp$ and small $B$).
Indeed, we can relax the condition on the smallness of $B$, by reducing $t_{\rm p}$. This can be done without enhancing the 
non-adiabatic corrections by increasing the environment coupling (or the classical noise power).

Finally, we note that Eq.~(\ref{Eq:NA1-non-degen}) indicate that there is now a small leading-order non-adiabatic correction to the phase, $\Phi_{\rm NA}^{(1)}$.  
However this can be neglected under the above condition that $B \ll \om$, because it is much smaller than
$\Phi_{\rm NA}^{(2)}$, that we treated earlier in this article.
There is also a even smaller modification of the Berry phase,
however since it is so much smaller than 
$\Phi_{\rm NA}^{(2)}$ it can also be neglected (indeed it would be
almost impossible to observe).

\section{Towards experimental observation}

The effects that we discuss here should be observable
in experimental set-ups similar to those used to observe conventional
Berry phases, such as the superconducting qubits experiments of 
Ref.~[\onlinecite{Wallraff-expt}] or the cold-neutron experiments of Ref.~[\onlinecite{Rauch-expt}].  

The effects induced by strong non-Markovian classical noise should be relatively straightforward to observe.  If one is to able to drive the
three-components of the spin with arbitrary signals, then one
can drive them with a man-made noise signal.  
One could follow Ref.~[\onlinecite{Rauch-expt}], and construct this noise-signal
on a computer, thereby controlling all aspects of the noise.
Importantly one can then choose the components of the noise
to be correlated in the manner that indicates that the noise has a natural axis,
which varies with time around a closed loop.
This would be sufficient to show all the effects that we discuss associated with classical noise; noise-generated Berry phase, absence of the leading non-adiabatic correction, dephasing which goes to zero as we make the 
experiment time longer.

To see the same effects for strong coupling to a non-Markovian
quantum environment at low temperature will be much more challenging.
At present, we cannot think of an example of an experimental system in which this could be performed.  However a pre-requisite for a quantum computer
is that one has a large number of qubits with controllable couplings.
To observe the environment-induced Berry phase that we discuss, 
one probably only requires a few tens of qubits (much less than required for quantum computing).
One qubit would play the role of the system,
and would be tuned to it degeneracy point.
The other qubits would play the role of the environment,
and would adjusted to have a variety of finite excitation frequencies.  
Then the coupling between the system qubit and environment qubit
would be varied in time as in Eq.~(\ref{Eq:H}).
With such a quantum environment, one can enter the regime where dephasing scales like $t_{\rm p}^{-3}$ rather than $t_{\rm p}^{-1}$
(because the  $t_{\rm p}^{-1}$-term has an exponentially small prefactor). 
However apart from this additional suppression of dephasing (which is already small
for classical noise), this environment induces the same effects as classical noise.
Thus, since classical noise is easier to implement experimentally, we see no
significant reason to try to implement the quantum environment.

\section{Concluding remark.} 

This work --- on the physics of the orthogonality catastrophe 
and the associated strong renormalization of spin dynamics by a highly non-Markovian environment
--- arrives at two intriguing conclusions. 
The first is that it appears that 
adding coupling to an environment (of the right type) may make it {\it easier} to accurately measure Berry phases, when naively one would say an environment always suppresses such phase information.
The second is that even {\it classical} non-Markovian noise induces the strong renormalization of spin dynamics necessary to generate a Berry phase; even though such renormalization is based on the
intrinsically {\it quantum} effect of vanishing overlaps between environment states.  This means that the classical limit of this quantum effect, must correspond to a classical effect.  Unfortunately the nature of the 
calculation in this article (a formal mapping from the quantum problem to
the  classical one) does not help us understand this classical effect.
In future work we plan to study the classical problem directly,
and find a simple picture of the classical effect that plays a role analogous to the $T \to \infty$ limit of the orthogonality catastrophe.

Finally, we recall that our results for a quantum environment are valid for arbitrary temperatures
if the environment is made up of harmonic oscillators.  However, as discussed in Section~\ref{sect:beyond-oscillators}, the low temperature limit of these results apply for an arbitrary environment (a bath of two-level systems, etc).

\section{Acknowledgments}
My thanks go to  R.~Leone and L.~L\'evy for inspiring this work, 
to M.~Berry and P.~Bruno for insights into geometric phases, 
to S.~Florens for insights into spin-bosons models and adiabatic renormalization,
and to R.~Englman for insights that led me to the polaron literature. 
Finally, I thank T.~Ziman for his general advice on RG and his ongoing support 
throughout this work.

\appendix

\section{The Aharonov-Anandan phase}
\label{append:AA}

There exists another geometric phase which has no non-adiabatic 
corrections; the Aharonov-Anandan (AA) phase \cite{AA}.
Thus it is natural to ask why we are working so hard to create a Berry phase
with suppressed non-adiabatic corrections, when the AA phase already exists.
Our response to this question is that the nature of the Berry phase makes it much more 
easier to calculate in most experimental situations than the AA phase.
Here we explain what we mean by this.

The Berry phase is given by the solid-angle enclosed by the parameters of the Hamiltonian,
which are directly controlled by the experimenter.  Thus for any given time-dependence
of those parameters the Berry phase can be calculated with simple trigonometry.

In contrast, the AA phase is half the solid-angle enclosed by those quantum states
that return to themselves  \cite{AA}.  
While the concept of the AA phase is certainly mathematically beautiful, 
let us consider for a moment the amount
of work the experimenters have to do to calculate the AA phase for a given time-dependence
of the parameters of  Hamiltonian.  They must first completely solve the time-dependent Schr\"odinger
equation for evolution from time zero to time $t_{\rm p}$ to find those states which
return to themselves.  For an arbitrary time-dependence of the parameters of the Hamiltonian, this cannot be done analytically, so they must numerically simulate the quantum evolution to find these states.
Having done this, they must then find the solid-angle enclosed by these states.  This involves recording
their behaviour for all time from zero to $t_{\rm p}$.
For arbitrary time-dependence of the Hamiltonian, 
this typically involves re-running the numerical 
simulation to get the time-dependence of those states.
Only then can one use trigonometry to calculate the solid angle enclosed
to extract the AA phase.

Of course, there are certain simple 
cases in which the AA phase can be calculated analytically \cite{AA-qubits}; 
e.g., spin-half in a field which is static from time zero to time $t'$, then instantaneously
changes to a new direction, and then remains static for times from $t'$ to $t_{\rm p}$.  However these cases are exceptional.

Thus one could say that the noise-generated Berry phase, that we present here,
fills a gap between the conventional Berry phase and the AA phase.
It is as easy to calculate as a conventional Berry phase,
while having much smaller non-adiabatic corrections.

\section{Tranformation to the basis of shifted bath-modes}
\label{append:trans-2-shifted}

We label the eigenstates of $\hat{\sigma}_z$ with $s=\pm 1\equiv \up,\dn$,
then $\sigma_x = \sum_s |-s\rangle \langle s|$.
If there were only one bath mode (the $j$th mode), then 
we  would insert a resolution of unity of the form
$1=\sum_m |m_j^{s}\rangle \langle m_j^{s}|$ to the right
and another of the form $1=\sum_{n} |n_j^{-s}\rangle \langle n_j^{-s}|$ to the left,
giving 
\begin{widetext}
\begin{eqnarray}
\fl \hat{\sigma}_x 
&=& \sum_s |-s\rangle \langle s|\ \sum_{m_j}
\Big[ |m_j^{-s}\rangle \langle m_j^{-s}|m_j^{s}\rangle \langle m_j^{s}| \,+\, 
|(m-1)_j^{-s}\rangle\langle (m-1)_j^{-s}|m_j^{s}\rangle\langle m_j^{s}|
\nonumber \\
\fl & &\qquad \qquad \qquad+ \, 
|(m+1)_j^{-s}\rangle\langle (m+1)_j^{-s}|m_j^{s}\rangle\langle m_j^{s}|
+ \cdots \Big] 
\nonumber \\
\fl &=& \hat{\sigma}'_{x(j)}
\Big[ 1- \al_j^2(\hat{a}_j\hat{a}_j^\dagger +\hat{a}_j^\dagger\hat{a}_j)/2
-\al_j \hat{\sigma}'_{z(j)} (\hat{a}_j^\dagger-\hat{a}_j)
+ {\cal O}[\al_j^2\hat{a}_j^{\dagger 2}, \al_j^2\hat{a}_j^2,\al_j^4] \Big], \qquad
\end{eqnarray}
where we define
$\hat{a}_j^{\prime \dagger} = 
(m+1)^{1/2} \ \big|(m+1)_j^s \big\rangle \, \big\langle m_j^s \big|$,
$\hat{a}_j^{\prime} = m^{1/2}\ \big|(m-1)_j^s \big\rangle \,\big\langle m_j^s \big|$,
and 
\begin{eqnarray}
\hat{\sigma}'_{x(j)} 
&=& \sum_s |-s\rangle \langle s| \, |m_j^{-s}\rangle \langle m_j^{s}|,
\qquad \hat{\sigma}'_{z(j)} 
= \sum_s s\, |s\rangle \langle s| \, |m_j^{s}\rangle \langle m_j^{s}|.
\end{eqnarray}
\end{widetext}
Note that  we assume that the environment is extremely large, and so the coupling to any 
given environment mode is tiny enough that we can keep only lowest (second) order terms in $\al_j^2$. Since we trace over the environment at the end,  
terms of ${\cal O}[\al_j^2\hat{a}_j^{\dagger 2}]$ or of ${\cal O}[\al_j^2\hat{a}_j^2]$ can be neglected, 
because they can only lead to contributions of ${\cal O}[\al_j^4]$.

There are $N$ bath modes, so we insert $N$ resolutions 
of unity on the left and another $N$ on the right. 
We then define 
\begin{eqnarray}
\hat{\sigma}'_{x} 
&=& \sum_s |\!-\!s\rangle \langle s| \, 
\prod_{j=1}^N |m_j^{-s}\rangle \langle m_j^{s}|
 \, =\, \left(\begin{array}{cc} 0 & 1 \\ 1 & 0 \end{array} \right),
\nonumber \\
\hat{\sigma}'_{y} 
&=& \sum_s \rmi s |\!-\!s\rangle \langle s| \, 
\prod_{j=1}^N |m_j^{-s}\rangle \langle m_j^{s}|
 \, =\, \left(\begin{array}{cc} 0 & -\rmi \\ \rmi & 0 \end{array} \right),
\nonumber \\
\hat{\sigma}'_{z} 
&=& \sum_s s\, |s\rangle \langle s| \, 
\prod_{j=1}^N |m_j^{s}\rangle \langle m_j^{s}|
 \, =\, \left(\begin{array}{cc} 1 & 0 \\ 0 & -1 \end{array} \right),
\label{Eq:primed-sigmas}
\end{eqnarray}
where the second equality (equating the operators 
to the usual Pauli matrices) is only valid if we are working in the basis
of shifted bath-modes.
Following the same logic
as for one bath mode, one gets
\begin{eqnarray}
\hat{\sigma}_x 
&=& \hat{\sigma}'_x \exp[-\hat{F}] 
\prod_j 
\Big[1-\al_j \hat{\sigma}'_z (\hat{a}_j^\dagger-\hat{a}_j) \Big], \quad
\label{Eq:sigma_x-intermediate}
\end{eqnarray}
where we use the fact that $\al_j$ is very small (while $N$ is very large),
to write $\prod_j \big(1-\al_j^2
(\hat{a}_j\hat{a}_j^\dagger +\hat{a}_j^\dagger\hat{a}_j)/2\big) 
\simeq 
\prod_j \exp\big[- \al_j^2
(\hat{a}_j\hat{a}_j^\dagger +\hat{a}_j^\dagger\hat{a}_j)/2\big] 
= \exp[-\hat{F}]$.
Further, the assumption of very large $N$ means that $\langle \hat{F}\rangle$ remains unchanged
during the evolution.
Thus we replace $\hat{F}$ by $F= \langle \hat{F}\rangle$ given in Eq.~(\ref{Eq:F-any-env}).
For an environment of harmonic oscillators at temperature $T$,
$\langle\hat{a}_j^\dagger\hat{a}_j\rangle = \exp[-\beta\Om_j] /(1- \exp[-\beta\Om_j])$
where $\beta=(k_{\rm B}T)^{-1}$, then
we get the result below Eq.~(\ref{Eq:F-any-env}).

Eq.~(\ref{Eq:sigma_x-intermediate}) is easier to handle if we extract 
the spin-operators from the product as follows.
We note that all terms with even (odd) numbers of  $\al_j$s, have an even (odd) power of $\hat{\sigma}'_z$. Thus as $\hat{\sigma}^{\prime 2}_z=1$, terms with even numbers of  $\al_j$s
contain no spin-operator, while terms with odd numbers of  $\al_j$s
contain a single $\hat{\sigma}'_z$ operator.
We separate terms with even and odd numbers of $\al_j$s, by noting that
all even terms in $\prod_j [1-x_j]$ are given by 
$\half \big(\prod_j [1-x_j] +\prod_j [1+x_j]\big)$, while all
odd terms are given by the difference of these two terms.
Thus we get 
\begin{eqnarray}
\hat{\sigma}_x&=&
\hat{\sigma}'_x\e^{-F} + \hat{\sigma}'_x{\cal Q}_{\rm even}  
+\hat{\sigma}'_y{\cal Q}_{\rm odd},
\label{Eq:sigma_x}
\\
\hat{\sigma}_y
&=&  \hat{\sigma}'_y \e^{-F}
+ \hat{\sigma}'_y {\cal Q}_{\rm even}
- \hat{\sigma}'_x {\cal Q}_{\rm odd},
\label{Eq:sigma_y}
\\
\hat{\sigma}_z &=& \hat{\sigma}'_z,
\label{Eq:sigma_z}
\end{eqnarray}
where ${\cal Q}_{\rm even,odd}$ are the Hermitian environment operators
given in Eq.~(\ref{Eq:Q}).
Writing the ``universe'' Hamiltonian, ${\cal H}_{\rm sys\& env}$,
in terms of these results, we arrive at Eq.~(\ref{Eq:Hprime}).

\section{Evaluating traces over the environment}
\label{append:env-trace}

The  Born approximation of $\bSigma$ 
in Eq.~(\ref{Eq:exact-master}) is most naturally written 
with the system and environment operators in the interaction picture,
thus we define
$
\hat{\sigma}'_{x,y}(-\tau) = \e^{-\rmi {\cal H}'_{\rm sys}\tau} 
\hat{\sigma}'_{x,y} \e^{\rmi {\cal H}'_{\rm env}\tau}$ and 
$
{\cal Q}_{\rm even,odd}(-\tau) = \e^{-\rmi {\cal H}'_{\rm env}\tau} 
{\cal Q}_{\rm even,odd} \e^{\rmi {\cal H}'_{\rm env}\tau}$.
Then ${\cal Q}_{\rm even,odd}(t)$ are given by Eq.~(\ref{Eq:Q}) 
with $\hat{a}^{\prime\dagger}_j \to \e^{\rmi \Om_j t}\hat{a}^{\prime\dagger}_j$ and
$\hat{a}'_j \to \e^{-\rmi \Om_j t}\hat{a}^{\prime\dagger}_j$.
We define the symmetric and  asymmetric noise functions
\begin{eqnarray}
A_{\rm even}^{\rm sym}
&=& \half \tr_{\rm env}
\Big[\rho_{\rm env}^{\rm ini} 
\big[{\cal Q}_{\rm even}(0), {\cal Q}_{\rm even}(-\tau)\big]_+ \Big],
\nonumber \\
A_{\rm even}^{\rm asym} 
&=& -\rmi \half \tr_{\rm env}
\Big[\rho_{\rm env}^{\rm ini} 
\big[{\cal Q}_{\rm even}(0), {\cal Q}_{\rm even}(-\tau)\big]_- \Big], \qquad 
\label{Eq:Asym-asym}
\end{eqnarray}
where $[{\cal X},{\cal Y}]_\pm = {\cal X}{\cal Y}\pm {\cal Y}{\cal X}$.
We define $A_{\rm odd}^{\rm sym}$ and $A_{\rm odd}^{\rm asym}$ in a similar manner
with ``even'' $\to$ ``odd''. 
We note that $A_{\rm even,odd}^{\rm sym,asym}$ are all defined to be real.
To find   $A_{\rm even,odd}^{\rm sym,asym}$,  we need
\begin{widetext}
\begin{eqnarray}
\tr_{\rm env}\big[\rho^{\rm env}_0 {\cal Q}_{\rm even}(0) 
{\cal Q}_{\rm even}(-\tau)\big]
 = \e^{-2F}
{\rm tr}_{\rm env}\big[\rho^{\rm env}_0 
\big({\cal R}_+(0) +{\cal R}_-(0)-1\big)
\big({\cal R}_+(-\tau) +{\cal R}_-(-\tau)-1\big)\big],
\end{eqnarray}
where
${\cal R}_\pm(-\tau) = \half \prod_{j=1}^N 
\big(1 \mp \al_j\hat{a}_j^{\prime\dagger}\e^{-\rmi \Om_j\tau}\pm 
\al_j\hat{a}_j^\prime\e^{\rmi \Om_j\tau}\big)$.
For diagonal $\rho^{\rm env}_0$ (i.e. a thermal state), 
\begin{eqnarray}
\tr_{{\rm env}; j} 
\Big[\rho_0^{{\rm env};j}
\big(1+\al_j\hat{a}_j^{\prime}
-\al_j\hat{a}_j^{\prime\dagger}
\big)
\big(1-\al_j\hat{a}_j^{\prime\dagger}\e^{-\rmi \Om_j\tau}
+\al_j\hat{a}_j^{\prime}\e^{\rmi \Om_j\tau}\big)
\Big]
&=& 1- \al_j^2\nu^{(2)}_j\e^{-\rmi \Om_j\tau} 
-\al_j^2\nu^{(1)}_j\e^{\rmi \Om_j\tau},
\end{eqnarray}
\end{widetext}
where we define 
$\nu^{(1)}_j=\langle\hat{a}_j^{\prime\dagger}\hat{a}_j^{\prime}\rangle$
and
$\nu^{(2)}_j=\langle\hat{a}_j^{\prime}\hat{a}_j^{\prime\dagger}\rangle$.
Thus for the trace over all $N$ environment modes we have 
\begin{eqnarray}
\tr_{\rm env}\big[\rho^{\rm env}_0 {\cal Q}_{\rm even}(0) 
{\cal Q}_{\rm even}(-\tau)\big]
= \e^{-2F}\big( \cosh [2f(\tau)]-1 \big),
\nonumber \\
\label{Eq:trace-even-even}
\end{eqnarray}
where 
$f(\tau) = \half \sum_{j=1}^N \al_j^2 \big(
\langle\hat{a}_j^{\prime}\hat{a}_j^{\prime\dagger}\rangle
\e^{-\rmi \Om_j\tau} +
\langle\hat{a}_j^{\prime\dagger}\hat{a}_j^{\prime}\rangle 
\e^{\rmi \Om_j\tau}\big)
$
and we recall that $\al_j = K_j/\Om_j$.
This definition of $f(\tau)$ is such that $f(0)$ equals the Franck-Condon factor, $F$, in Eq.~(\ref{Eq:F-any-env}).
To get the last line of Eq.~(\ref{Eq:trace-even-even}), 
we note that
$|\al_j(0)\al_j(-\tau)| \leq |\al_j|^2\ll 1$ 
and hence use the approximation 
$\prod_j(1-f_j) \simeq \prod_j \exp[-f_j] =
\exp \big[-\sum_j f_j \big]$.
We evaluate other traces in the same manner,  we need
\begin{eqnarray}
\tr_{\rm env}\big[\rho^{\rm env}_0 {\cal Q}_{\rm odd}(0) 
{\cal Q}_{\rm odd}(-\tau)\big]
&=& \e^{-2F}\sinh [2f(\tau)], \qquad
\\
\tr_{\rm env}\big[\rho^{\rm env}_0 {\cal Q}_{\rm odd}(0) 
{\cal Q}_{\rm even}(-\tau)\big]
&=& 0.
\end{eqnarray}
The second result comes from the fact that 
${\cal Q}_{\rm even}$ and ${\cal Q}_{\rm odd}$ 
respectively cause the transition
of an even and odd number of environment modes. Thus their product
involves the transition of a odd number of modes, so the trace is zero.

Using the definition of the dimensionless $J$-function below Eq.~(\ref{Eq:chi_n}), we  write
\begin{eqnarray}
f(\tau)= {1 \over 2\pi} \int_0^\infty {\rmd x \ j(x)\over x^2} \ {\e^{-\rmi \Om_{\rm m}\tau x} +\e^{\rmi \Om_{\rm m}\tau x -\beta \Om_{\rm m}x} \over 1-\e^{-\beta \Om_{\rm m}x}  },
\end{eqnarray}
and then approximate $f(\tau)$ by its expansion in $\tau$,
\begin{eqnarray}
\e^{f(\tau)} &\simeq& \exp\Big[\Amp \big(\chi_0 - \rmi \chi_1\Om_{\rm m}\tau
-\half (\Om_{\rm m}\tau)^2
\nonumber \\
& & \qquad \qquad \qquad + {\cal O}[\chi_3 (\Om_{\rm m}\tau)^3 ] \big) \Big],
\label{Eq:saddle-point}
\end{eqnarray}
where $\chi_n$ is given in Eq.~(\ref{Eq:chi_n}), with $\chi_2=1$ by construction.
This approximation is natural for a non-Markovian environment, since it has $\Amp \gg 1$
(see Section~\ref{sect:Hs-qu}), since we can assume that when we integrate
over $\tau$, we do so in a saddle-point approximation.
So long as  $\chi_3 \ll \Amp^{1/2}$, 
the quadratic term in the above exponents ensures that $\e^{f(\tau)}$
decays to be much less than $\e^{f(0)}$ long before the 
$(\Om_{\rm m}\tau)^3$-term becomes relevant 
(which is why we were able to drop it from the exponent).
For an environment without low-frequency modes $\chi_3 \sim 1$,
thus $\Amp \gg 1$, 
we find that  $A_{\rm even,odd}^{\rm sym,asym}$ are given by Eq.~(\ref{Eq:all-As-strong}).

For classical noise  $\chi_3=0$, so the conditions under which we can apply 
Eq.~(\ref{Eq:saddle-point}) are a little different.
Since condition that  $\Amp \gg \chi_3^2$ is trivially fulfilled, saddle-point approximation is applicable under the condition that we can neglect the quartic term in the exponent, $\Amp \chi_4 (\Om_{\rm m}\tau)^4$.  It can be neglected if $\chi_4 \ll \Amp$. 
For noise without low-frequency modes, $\chi_4 \sim 1$, so the condition is satisfied when 
$\Amp \gg 1$.

\vskip 4mm


\begin{thebibliography}{9}
\bibitem{footnote:units}
We gives all energies and B-fields in units of inverse time.
\bibitem{Anderson-footnote}
In this context, the electron tunnels ``slowly'' from a state in one metal electrode
to a state in the other.
The tunnelling electron repels all the other electrons in the electrodes 
(where they have ``fast'' dynamics), thus these electrons
will act as a  ``cloud'' that can strongly renormalize the tunnelling rate.
\end{thebibliography}
\end{document}